\documentclass[aps,prl,reprint,groupedaddress,longbibliography,preprintnumbers]{revtex4-2}

\usepackage[utf8]{inputenc}
\usepackage{amsmath,amssymb}
\usepackage{amsfonts}
\usepackage{slashed}
\usepackage{xcolor}
\usepackage{tensor}

\usepackage[pdftex,hypertexnames=false]{hyperref}
\usepackage{bbm}
\hypersetup{ 
pdfstartview=FitV,
colorlinks=true, 
linkcolor=black, 
citecolor=blue, 
filecolor=black, 
urlcolor=magenta,
}
\usepackage{graphicx}
\usepackage[caption=false]{subfig}
\usepackage{ulem}

\begin{document}
	\preprint{CPHT-RR064.122022}
	\title{Leading order magnetic field dependence of conductivities in anomalous hydrodynamics}
	\author{Andrea Amoretti$^{\, 1,2}$, Daniel K. Brattan$^{\, 1,3}$, Luca Martinoia$^{\, 1,2}$ and Ioannis Matthaiakakis$^{\, 1,2}$.}
	\affiliation{\textbf{1} Dipartimento di Fisica, Universit\`a di Genova,
		via Dodecaneso 33, I-16146, Genova, Italy,
	}
	\affiliation{\textbf{2} I.N.F.N. - Sezione di Genova, via Dodecaneso 33, I-16146, Genova, Italy,}
	\affiliation{\textbf{3} CPHT, CNRS, École polytechnique, Institut Polytechnique de Paris, 91120 Palaiseau, France.
}
	
	\email{andrea.amoretti@ge.infn.it}
	\email{danny.brattan@gmail.com}
	\email{luca.martinoia@ge.infn.it}
	\email{\\ioannis.matthaiakakis@edu.unige.it}
	\begin{abstract}
		{\noindent  We show that  literature results claimed for the magnetic field dependence of the longitudinal conductivity in anomalous first-order hydrodynamics are frame dependent at this derivative order. In particular, we focus on $(3+1)$-dimensional hydrodynamics in the presence of a constant ${\cal O}(\partial)$ magnetic field with a $U(1)$ chiral anomaly and demonstrate that, for constitutive relations up to and including order one in derivatives, the anomaly drops out of the longitudinal conductivity. In particular, magnetic field dependent terms that were previously found in the literature only enter the non-zero frequency thermoelectric conductivities through explicitly frame dependent pieces indicating that they are not physical. This issue can be avoided entirely by incorporating the magnetic field into the fluid's equilibrium state.}
	\end{abstract}
	
	\maketitle

	{\noindent Hydrodynamics is an effective theory of thermalized matter at length and time scales far larger than any dynamical scale in the system. As any effective field theory, hydrodynamics then takes the form of a derivative expansion that truncates to the order most relevant to the situation one wishes to analyze. This truncation, while making hydrodynamics more tractable, also ushers in an ambiguity in its precise definition; the choice of frame. In a nutshell, the choice of frame allows us to choose how our system deviates from its global equilibrium configuration to a local one given by the solution of the equations of hydrodynamics. In principle the number of such choices is infinite, but only a handful of those are ever used in the literature; the Landau frame, the Eckart frame, the thermodynamic frame and the BDNK frame \cite{Bemfica:2017wps,Kovtun:2019hdm}. For example, the Landau frame is defined in terms of the energy-momentum tensor $T^{\mu\nu}$ and current $J^\mu$ of a fluid via the conditions
\begin{equation}
\label{Eq:LandauFrameDef}
T^{\mu\nu}u_\nu = -\epsilon u^\mu ~~,~~ J^\mu u_\mu = -n~,
\end{equation}	
with $u_\mu$ the velocity profile of the fluid and $\epsilon~,n$ its equilibrium energy and charge density respectively. On the other hand, the thermodynamic frames \footnote{Equation \eqref{Eq:ThermoFrameDefinition} defines a frame only up to terms that vanish at hydrostaticity. Hence there is a large class of thermodynamic frames which differ by terms that vanish upon choosing solutions satisfying \eqref{Eq:ThermoFrameDefinition}.} are defined in terms of a timelike Killing vector $V^\mu$ as 
\begin{align}
		\label{Eq:ThermoFrameDefinition}
		& T = \frac{T_{0}}{\sqrt{-V^2}} \; ,  \; \;  \mu = \frac{A_{\mu} V^{\mu} + \Lambda_{V}}{\sqrt{-V^2}} \; , \; \;  u^{\mu} = \frac{V^{\mu}}{\sqrt{-V^2}} \; , 
\end{align}
where $T$ is the fluid's temperature with $T_{0}$ a normalisation constant and $\mu$ the fluid's chemical potential with $\Lambda_{V}$ a gauge parameter that maintains the gauge invariance of $\mu$. The gauge field $A_\mu$ is the source of $J^\mu$.}

	{\ Which frame one decides to use is a matter of convention, but the choice strongly affects the predicted non-hydrodynamic behaviour of physical observables and even the causal properties of hydrodynamics itself. For example, the Landau frame \eqref{Eq:LandauFrameDef} fixes the energy and charge density of the fluid to be the equilibrium ones, while the thermodynamic frame \eqref{Eq:ThermoFrameDefinition} makes sure the temperature and chemical potential are given by their canonical definitions in equilibrium. That these frames may give seemingly different physics is discussed in more detail in a recent paper \cite{kovtun:onequilibriumtemperature}, where it was argued that in hydrostatic configurations on curved backgrounds some potential definitions of the temperature, which differ from the canonical choice by derivatives of the metric, can have undesirable properties. Namely, the temperature is not related to the derivative of the entropy with respect to the energy i.e. $1/T \neq dS/dE$. }
	
{ Note that while frame-dependence is ubiquitous in theories of hydrodynamics, it should not be found in expressions of directly observable quantities, such as the fluid's thermoelectric conductivities. The reason behind this is that frame-dependence is an artifact of our inability to calculate directly obsevable quantities without truncating the hydrodynamic derivative expansion. The experimentalist who measures these quantities, however, does not have to make such a truncation; they measure the ``fully resummed'' version of that quantity. Therefore all theoretical predictions for directly observable quantities must be frame-independent, otherwise no comparison to experiment can be made.}
	
	{\ In this paper we consider hydrodynamics in $(3+1)$-spacetime dimensions exhibiting a $U(1)_A$ axial anomaly coupled to an order one in derivatives, ${\cal O}(\partial)$, magnetic field. We calculate the non-zero frequency, magnetic field dependent, thermoelectric conductivities of such a fluid and find that---at order one in derivatives---the anomaly only enters through frame dependent terms. This is because, when care is taken with derivative counting, we find that the anomaly-dependent terms lie beyond the ${\cal O}(\partial)$ hydrodynamic regime. If we wish to observe non-trivial effects from the anomaly at ${\cal O}(\partial)$ in the constitutive relations (the order where incoherent conductivities first appear) without resorting to ${\cal O}(\partial^2)$ hydrodynamics, we must work with order zero magnetic fields. This implies, however, that we can never reach the Landau frame.}
	
	{\ }
	
	\textit{The standard approach to anomalous hydrodynamics. }{\noindent In general, mixed axial-gravitational anomalies can be present in addition to the gauge anomaly. Moreover, for physically relevant models one may wish to have an anomalous and a non-anomalous current such as in $U(1)_{A} \times U(1)_{V}$ models. As our results  readily generalize to these more involved cases, here we focus on the simple $U(1)_A$ model in $(3+1)$-spacetime dimensions with just one anomalous axial current $J^\mu$ \cite{landsteiner:negativemagnetoresistivity,abbasi:magnetotransportanomalous,son:hydrodynamicstriangle}. In what follows, we review the standard construction of these anomalous fluids, discuss their frame transformation properties and derive the corresponding thermoelectric conductivities.}
	
	{\ The equations of motion for the (anomalous) covariant current and the stress energy tensor we consider are
\begin{subequations}
\label{Eq:HydroEqns}
\begin{align}
    \partial_\mu T^{\mu\nu}&=F^{\nu\lambda}J_\lambda \; , \\
    \partial_\mu J^\mu&= c E^{\mu} B_{\mu} \; . 
\end{align}
\end{subequations}
The covariant axial electric and magnetic field are defined via the Maxwell field strength $F_{\mu\nu} = \partial_\mu A_\nu - \partial_\nu A_\mu$ via $E^\mu = F^{\mu\nu}u_\nu$ and $B^\mu=\frac{1}{2}\varepsilon^{\mu\nu\rho\sigma}u_\nu F_{\rho\sigma}$. Both $E^\mu$ and $B^\mu$ are considered $\mathcal{O}(\partial)$. The constant $c$ is the anomaly coefficient.}

	{\ To consider the effective hydrodynamic theory associated with \eqref{Eq:HydroEqns}, we work at non-zero temperature $T$ and axial chemical potential $\mu$. The axial chemical potential is distinct from the usual one, because it is odd under spatial parity reversal. Similarly for the axial charge current.}

	\textit{The constitutive relations.}
{The constitutive relations for an anomalous fluid to first order in derivatives are well known (see e.g. \cite{Nair:2011mk,Monteiro:2014wsa,landsteiner:notesanomaly}). We record them here for ease of reference,
\begin{subequations}
\label{Eq:Constitutive}
\begin{align}
    T^{\mu\nu}=\,&\epsilon u^\mu u^\nu+p\Delta^{\mu\nu} +\xi_B^\epsilon\left(u^\mu B^\nu+u^\nu B^\mu\right) \nonumber \\
    & +\xi_\Omega^\epsilon\left(u^\mu \Omega^\nu+u^\nu \Omega^\mu\right)  -\eta\Delta^{\mu\alpha}\Delta^{\nu\beta}\sigma_{\alpha\beta} \nonumber \\
    & -\zeta\Delta^{\mu\nu}\partial_\alpha u^\alpha + \mathcal{O}(\partial^2) \\
    J^\mu=\,&n u^\mu+\sigma_0\Delta^{\mu\nu}\left(E_\nu-T\partial_\nu\frac{\mu}{T}\right)+ \xi_{\Omega} \Omega^{\mu} + \xi_B B^\mu \nonumber \\
    	      & + \mathcal{O}(\partial^2) \; ,
\end{align}
\end{subequations}
where $\Delta^{\mu\nu}=\eta^{\mu\nu}+u^\mu u^\nu$ is the projector orthogonal to the four-velocity, $\sigma_{\mu\nu}$ the shear tensor, $\eta$ and $\zeta$ are respectively the shear and bulk viscosity, while $\sigma_0$ is the conductivity and $\Omega^\mu=\varepsilon^{\mu\nu\alpha\beta}u_\nu\partial_\alpha u_\beta$ is the vorticity. The quantities $\xi$ are the dissipationless anomalous transport coefficients. Their expression in terms of the hydrodynamic variables $\mu$ and $T$ is completely fixed by the anomaly and the choice of hydrodynamic frame \cite{Banerjee:2012cr}. For example, the Landau frame \eqref{Eq:LandauFrameDef} can be reached when $\xi^{\epsilon}_{B} =\xi^{\epsilon}_{\Omega}=0$ since
	\begin{subequations}
	\begin{eqnarray}
		T^{\mu \nu} u_{\nu} &=& - \left( \epsilon u^{\mu} + \xi^{\epsilon}_{B} B^{\mu}  + \xi^{\epsilon}_{\Omega} \Omega^{\mu} \right) \; , \\
		J^{\mu} u_{\mu} &=& - n \; .
	\end{eqnarray}
	\end{subequations}
}

{\ Given one frame, we can move to a different one by redefining the velocity \footnote{Frame transformations of $T$ and $\mu$ are also possible, but they do not affect our results for the thermoelectric conductivities.}. We are particularly interested in frame transformations of the kind
	\begin{subequations}
	\label{Eq:GenericFrameChange}
	\begin{equation}
		u^{\mu} \rightarrow u^{\mu} + f_{B}(\mu,T) B^{\mu} + f_{\Omega}(\mu,T) \Omega^{\mu}
	\end{equation}
which shift the anomalous transport coefficients as
	\begin{eqnarray}
		\xi^{\epsilon}_{B,\Omega} &\rightarrow& \xi^{\epsilon}_{B,\Omega} + (\epsilon + p) f_{B,\Omega}(\mu,T) \; , \\
		\xi_{B,\Omega} &\rightarrow& \xi_{B,\Omega} + n f_{B,\Omega}(\mu,T) \; ,
	\end{eqnarray}
	\end{subequations}
Note that such transformations are permitted since $B^{\mu}$ and $\Omega^{\mu}$ are ${\cal O}(\partial)$. When we refer to ``frame transformations'' in this paper we refer only to the subset  \eqref{Eq:GenericFrameChange}  displayed above unless explicitly stated otherwise. Said differently: The constitutive relations in \eqref{Eq:Constitutive} are given in the Landau frame when the anomaly coefficient or the magnetic field and vorticity are set to zero.}

\textit{Positivity of entropy production.}
{\ As suggested by their name, the anomalous coefficients are completely specified in terms of the anomaly. In this subsection, we review the approach presented in \cite{son:hydrodynamicstriangle} to compute their explicit expressions. The derivation begins with the entropy current $S^\mu$, which we take to be of the form
	\begin{eqnarray}
		\label{Eq:EntropyCurrent}
		S^{\mu} &=& \frac{1}{T} \left( p u^{\mu} - T^{\mu \nu} u_{\nu} - \mu J^{\mu} \right) + S^{\mu}_{\mathrm{eq.}}
	\end{eqnarray}
where $S^{\mu}_{\mathrm{eq.}}$ is the non-canonical entropy current, needed to ensure positivity of entropy production in the presence of an axial anomaly. In our case
	\begin{eqnarray}
		S^{\mu}_{\mathrm{eq.}} &=& \xi_{B}^{s} B^{\mu} + \xi_{\Omega}^{s} \Omega^{\mu} + \mathcal{O}(\partial^2) \; . 
	\end{eqnarray}
Taking the divergence of $S^\mu$ and employing the equations of motion \eqref{Eq:HydroEqns}  we find
	\begin{eqnarray}
			\label{Eq:Entropy}
			\partial_{\mu} S^{\mu}
		&=&	\frac{\zeta}{T} \theta^2 + \frac{\eta}{T} \sigma_{\mu \nu} \sigma^{\mu \nu} \nonumber \\
		&\;& + \sigma_{0} \left( E_{\mu} - T \partial_{\mu}^{\perp} \frac{\mu}{T} \right)  \left( E^{\mu} - T \partial^{\mu}_{\perp} \frac{\mu}{T} \right) \nonumber \\
		&\;& + \left(  \frac{\partial \xi^{s}_{B}}{\partial T} + \frac{\mu}{T} \frac{\partial \xi^{s}_{B}}{\partial \mu} - \frac{\xi^{\epsilon}_{B}}{T^2} \right) B^{\mu} \left( \partial_{\mu}^{\perp} T + T a_{\mu} \right) \nonumber \\
		&\;& +  \left(  \frac{\partial \xi^{s}_{\Omega}}{\partial T} + \frac{\mu}{T} \frac{\partial \xi^{s}_{\Omega}}{\partial \mu} - \frac{\xi^{\epsilon}_{\Omega}}{T^2} \right) \Omega^{\mu} \left( \partial_{\mu}^{\perp} T + T a_{\mu} \right)  \nonumber \\
		&\;& - \left( \frac{\partial \xi^{s}_{B}}{\partial \mu} - \frac{\xi_{B}}{T} \right) B^{\mu} \left(  E_{\mu}  - T \partial_{\mu}^{\perp} \left( \frac{\mu}{T} \right) \right) \nonumber \\
		&\;& - \left( \frac{\partial \xi^{s}_{\Omega}}{\partial \mu} - \frac{\xi_{\Omega}}{T} \right) \Omega^{\mu} \left(  E_{\mu} - T \partial_{\mu}^{\perp} \left( \frac{\mu}{T} \right) \right) \nonumber \\
		&\;& + \left( \xi^{s}_{B} - T \frac{\partial \xi^{s}_{B}}{\partial T} - \mu \frac{\partial \xi^{s}_{B}}{\partial \mu} \right) B^{\mu} a_{\mu} \nonumber \\
		&\;& + \left( 2 \xi^{s}_{\Omega} - T \frac{\partial \xi^{s}_{\Omega}}{\partial T} - \mu \frac{\partial \xi^{s}_{\Omega}}{\partial \mu} \right) \Omega^{\mu} a_{\mu} \nonumber \\
		&\;& + \left( \frac{\partial \xi^{s}_{B}}{\partial \mu} - \frac{c \mu}{T} \right) E^{\mu} B_{\mu} + \left( \frac{\partial \xi^{s}_{\Omega}}{\partial \mu}  - \xi^{s}_{B} \right) E^{\mu} \Omega_{\mu} \nonumber \\
		&\;& \geq 0 \; , 
	\end{eqnarray}
where we have used the Bianchi identity satisfied by $F_{\mu\nu}$ and defined $\partial_{\mu}^{\perp} = \Delta_{\mu \nu} \partial^{\nu}$. There are two types of constraint in \eqref{Eq:Entropy} - inequality type which require that $\sigma_{0}, \zeta, \eta \geq 0$ and equality type (the remainder), which cannot be made positive definite and thus must vanish identically on any solution to the hydrodynamic equations.}

{\ We can further use the ideal fluid equations of motion to find the relations
\begin{subequations}
	\begin{align}
		\partial_\mu\Omega^\mu&=-\frac{2}{\epsilon+p}\Omega^\mu\left(\partial_\mu p-nE_\mu\right)\ ,\\
		\partial_\mu B^\mu&=-2\Omega^\mu E_\mu+\frac{1}{\epsilon+p}\left(nE_\mu B^\mu-B^\mu\partial_\mu p\right)\ .
	\end{align}
\end{subequations}
Together with the entropy constraint, we then find the following set of equations for the $\xi$
\begin{subequations}
	\label{Eq:AnomalyEqns}
	\begin{align}
		&\partial_\mu\xi^s_\Omega-\frac{2\partial_\mu p}{\epsilon+p}\xi^s_\Omega-\xi_\Omega\partial_\mu\frac{\mu}{T}+\left(\frac{2\partial_\mu p}{\epsilon+p}-a_\mu-\partial_\mu\right)\xi^\epsilon_\Omega=0\ ,\\
		&\partial_\mu\xi^s_B-\frac{\partial_\mu p}{\epsilon+p}\xi^s_B-\xi_B\partial_\mu\frac{\mu}{T}+\left(\frac{\partial_\mu p}{\epsilon+p}-a_\mu-\partial_\mu\right)\xi^\epsilon_B=0\ ,\\
		&\frac{2n\xi^s_\Omega}{\epsilon+p}-2\xi^s_B+\frac{\xi_\Omega}{T}+2\xi^\epsilon_B-2\xi^\epsilon_\Omega \frac{n}{\epsilon+p}=0\ ,\\
		&\frac{n\xi^s_B}{\epsilon+p}+\frac{\xi_B}{T}-c\frac{\mu}{T}-\xi^\epsilon_B\frac{n}{\epsilon+p}=0 \ .
	\end{align}
\end{subequations}
}

{\ The set of equations \eqref{Eq:AnomalyEqns} is not closed as there are six unknown transport coefficients $\xi$ for only four equations. To explicitly solve for the anomalous transport coefficients some other constraint is required and this choice defines a specific frame. In particular the Landau frame condition $\xi^\epsilon_B=\xi^\epsilon_\Omega=0$ gives -- up to integration constants related to the mixed-gravitational anomaly and CPT violating terms \cite{neiman:relativistichydrodynamics,jensen:anomalyinflow,ammon:chiralhydrodynamics} -- the following transport coefficients,
\begin{subequations}
\label{Eq:XiLandau}
	\begin{align}
		\xi^\epsilon_B&=0	&	\xi_B&=c\left(\mu-\frac{1}{2}\frac{n\mu^2}{\epsilon+p}\right)\ ,\\
		\xi^\epsilon_\Omega&=0		&	\xi_\Omega&=c\left(\mu^2-\frac{2}{3}\frac{n\mu^3}{\epsilon+p}\right)\ .
	\end{align}
\end{subequations}
If however one wants to work in the thermodynamic frames \eqref{Eq:ThermoFrameDefinition}, the frame transformations \eqref{Eq:GenericFrameChange} can be used with
	\begin{eqnarray}
		f_{B} =  \frac{c \mu^2}{2 (\epsilon + p)} \; , \qquad f_{\Omega} =   \frac{c \mu^3}{3 (\epsilon + p)} \; . 
	\end{eqnarray}
This leads to the anomalous transport coefficients in the thermodynamic frame
\begin{subequations}\label{Eq:ThermodynamicFrame}
	\begin{align}
		\xi^\epsilon_B&=\frac{1}{2}c\mu^2 \; , 	&	\xi_B&=c\mu		\ ,\\
		\xi^\epsilon_\Omega&=\frac{1}{3}c\mu^3 \; , 	&	\xi_\Omega&=\frac{1}{2}c\mu^2	\ .
	\end{align}
\end{subequations}
}

\textit{Determining the conductivities.}
{\ Given our equations of motion \eqref{Eq:HydroEqns} and the constitutive relations \eqref{Eq:Constitutive} we can obtain the two-current Green's functions that describe the response of our system to hydrodynamic perturbations about equilibrium. In particular, we are interested in the two-current Green's function in the presence of a constant magnetic field so that we can extract the longitudinal magnetoconductivity of our fluid. To evaluate it, we linearise the hydrodynamic equations \eqref{Eq:HydroEqns} around an equilibrium configuration with constant temperature, chemical potential and magnetic field along $\hat{z}$ via $T=T_{0} + \delta T$, $\mu=\mu_{0}+\delta \mu$, $u^\mu=(1, \delta \vec{v})$, $F^{12}=B$ and $F^{0i}=\delta\mathbb{E}^i$. The equilibrium stress energy tensor and current are then given by
\begin{subequations}
\label{Eq:EquilibriumExpressions}
\begin{align}
    T^{\mu\nu}&=\begin{pmatrix}
    \epsilon    &   0   &   0   &   \xi_B^\epsilon B\\
    0   &   p   &   0   &   0\\
    0   &   0   &   p   &   0\\
    \xi_B^\epsilon B  &  0   &   0   &   p
    \end{pmatrix} + \mathcal{O}(\partial^2) \; ,  \\
    J^\mu&=\left(n,0,0,\xi_B B\right) + \mathcal{O}(\partial^2) \; .
\end{align}
\end{subequations}
}

	{\ Before solving for the perturbations, $(\delta T, \delta \mu,\delta v^{i})$, we note that we must be especially careful in our computation to keep track of the derivative order to which we are working \footnote{We thank Amos Yarom for making this point clear to us.} to ensure that any result is not an artefact of accidentally including spurious higher derivative terms. In particular, the hydrodynamic equations can schematically be written as 
		\begin{eqnarray}
			D X = S + \mathcal{O}(\partial^3) \; ,
		\end{eqnarray}
	where we are working with constitutive relations up to and including order one in derivatives. Here $X$ is the vector of our fluctuating fields $(\delta \mu,\delta T,\delta v^{i}$), $D$ is an operator of at least order one in derivatives and $S$ is a source term constructed from the hydrodynamic sources, the metric and gauge field and their derivatives. For constitutive relations given to first order in derivatives, the operator $D$ and the source term $S$ will contain terms up to and including second order.}
	
	{\ As an example of the above point, imagine introducing a formal ``derivative counting parameter'' $\varepsilon$ with $\omega,B \sim \varepsilon$. With only the longitudinal electric field turned on we find, after inverting $D$, the fluid velocity perturbation along the direction of the electric field reads
		\begin{eqnarray}
			\label{Eq:ExampleSoln}
			\left. \delta u_{z} \right|_{\vec{k}=\vec{0}} = \left[ \frac{- i n + \varepsilon^2 \alpha B^2 + \mathcal{O}(\varepsilon^3)}{(\epsilon + p) \omega \varepsilon + \mathcal{O}(\varepsilon^3)} \right] \delta \mathbb{E}_{z} \; . 
		\end{eqnarray}
A generic Green's function will then have the form
	\begin{eqnarray}\label{Eq:Green_function}
		G_{\mathrm{R}} \sim \frac{a(\omega,k) + \mathcal{O}(\varepsilon^2)}{b(\omega,k) + \mathcal{O}(\varepsilon^3)}
	\end{eqnarray}
where $a$ and $b$ are functions dependent on the precise correlator. The derivative counting in the numerator is a consequence of the fact that we know the currents only to ${\cal O}(\varepsilon)$. Consequently, the precise form of $\alpha$ in \eqref{Eq:ExampleSoln} is irrelevant and in principle the $B^2$ contribution to $G_{\mathrm{R}}$ must be dropped when considering an ${\cal O}(\partial)$ theory of hydrodynamics.}

{\ With these points made, taking the wavevector to zero, we find the following longitudinal electric conductivity,
		\begin{widetext}
	
	\begin{multline}\label{Eq:ACLongitudinalconductivity}
		\sigma(\omega)=\sigma_0+\frac{i n^2}{\omega w}+\frac{i B^2}{\omega w^2(\frac{\partial\epsilon}{\partial T} \frac{\partial n}{\partial\mu}-\frac{\partial\epsilon}{\partial\mu}\frac{\partial n}{\partial T})}\biggl[\Bigl(w\frac{\partial\xi_B}{\partial\mu}-n\frac{\partial\xi_B^\epsilon}{\partial\mu}\Bigr)\Bigl(w\Bigl(c\frac{\partial\epsilon}{\partial T}-\frac{\partial n}{\partial T}\xi_B\Bigr)-\frac{\partial\epsilon}{\partial T}n\xi_B+2\frac{\partial n}{\partial T}n\xi_B^\epsilon\Bigr)\\
		-\Bigl(w\frac{\partial\xi_B}{\partial T}-n\frac{\partial\xi_B^\epsilon}{\partial T}\Bigr)\Bigl(w\Bigl(c\frac{\partial\epsilon}{\partial\mu}-\frac{\partial n}{\partial\mu}\xi_B\Bigr)-\frac{\partial\epsilon}{\partial\mu}n\xi_B+2\frac{\partial n}{\partial\mu}n\xi_B^\epsilon\Bigr)\biggr]
	\end{multline}
		\end{widetext}
where we have expanded to $\mathcal{O}(B^2)$ in the magnetic field and we defined the enthalpy density $w=\epsilon+p$.  Note that the $\mathcal{O}(B^2)$ term is explicitly frame dependent because of its dependence on the $\xi$. For example in the Landau frame $\xi_B^\epsilon \equiv 0$, while in generic frames this is not true. Similar, frame dependent, results hold for the thermo-electric and thermal conductivities. 
As a consistency check, note that we reproduce the results of \cite{landsteiner:negativemagnetoresistivity} in the Landau frame.  The frame dependence of the conductivity is a sign that we are doing something wrong by keeping terms higher order than our constitutive relations allow us to do, as discussed above in \eqref{Eq:ExampleSoln} and \eqref{Eq:Green_function}. In particular, we emphasize that the magnetic field dependent term should not be considered physical since the $B^2$ (and higher) terms must be ignored, as they are higher order in derivatives than then one we, and most of the hydrodynamic literature, are working in.  Subsequently, the anomaly plays no role in the conductivity at this order in the derivative expansion.
 }
    
    {\ Moreover, if we naively take the longitudinal correlator without keeping track of the derivative order, we can readily find that the expressions in generic frames are not Onsager reciprocal; meaning $\alpha_{zz}\neq\bar{\alpha}_{zz}$. Furthermore the electric field and the chemical potential gradient do not appear as equivalent sources for the current via the Einstein combination $E_\nu - T\partial_\nu \mu/T$ in such cases.  It should also be noted that the frame dependence of \eqref{Eq:ACLongitudinalconductivity} is not a consequence of spurious UV poles \cite{kovtun:lectureshydrodynamic} - rather it is a consequence of spurious residues. Indeed, we have confirmed that the modes are frame independent (up to $\mathcal{O}(\partial^2)$). Similarly, the transverse correlators are independent of the anomalous transport terms and non-ambiguous when kept to the appropriate order.}
    
    {\ As a final note, an unusual feature of the anomalous hydrodynamics discussed above is the frame dependence of the equilibrium configuration \eqref{Eq:EquilibriumExpressions}. Typically, the equilibrium is frame independent and the result only depends on the transport coefficients, not on their derivatives \footnote{From a technical point of view, the derivatives appear because $\delta(\xi_BB^\mu)=\delta\xi_BB^\mu+\xi_B\delta B^\mu$, while usually in hydrodynamics the transport coefficients do not fluctuate as they multiply quantities that are zero in the background.}. Also, because the transport coefficients are not fixed by hydrodynamics, it is always possible to find a map that relates the results in two different frames. In the case of anomalous hydrodynamics however the equilibrium is frame dependent. Furthermore, the expressions of the transport coefficients in terms of $\mu$ and $T$ are entirely fixed in each frame, as in e.g. Eq.s~\eqref{Eq:XiLandau}  and \eqref{Eq:ThermodynamicFrame}. Therefore, there are no free parameters we can employ in order to relate results in different frames.}

	\textit{An unambiguous conductivity.}	
To resolve the frame ambiguity one could work at order two in derivatives \cite{abbasi:magnetotransportanomalous}, however this would only shift the problem to $\mathcal{O}(\partial^3)$ \footnote{The $B^2$ terms of the conductivities would be unambiguous for ${\cal O}(\partial^2)$, but including $B^4$ terms would result in frame dependent conductivities once again.}. To avoid the issue altogether, in order to study chiral fluids more easily and at larger values of $B$, we suggest to keep the magnetic field order zero in derivatives. By working with order zero in derivative magnetic fields we can bring anomalous effects to the ideal level of hydrodynamics where there are no frame redefinitions. If one then continues and determines constitutive relations at order one in derivatives then frame transformations of the form \eqref{Eq:GenericFrameChange} involving the magnetic field are not available because $B^{\mu}$ is no longer derivative order. The resultant expressions will not be in the Landau frame and in fact such a frame cannot be reached. Moreover, in this background we have an unambiguous global equilibrium because we cannot perform frame transformations containing $B^\mu$.

{\ A theory with ${\cal O}(\partial^0)$ magnetic field has already been studied in \cite{ammon:chiralhydrodynamics} and using their constitutive relations we can find expressions for the conductivities. The computation is exactly the same as performed in the standard approach: following the same notation as in \cite{ammon:chiralhydrodynamics} we decompose the stress energy tensor and current as
\begin{subequations}
\label{Eq:TensorCurrentDecomp}
\begin{align}
    T^{\mu\nu}&=\mathcal{E}u^\mu u^\nu+\mathcal{P}\Delta^{\mu\nu}+\mathcal{Q}^\mu u^\nu+\mathcal{Q}^\nu u^\mu+\mathcal{T}^{\mu\nu}\\
    J^\mu&=\mathcal{N}u^\mu+\mathcal{J}^\mu
\end{align}
\end{subequations}
and we focus on the ideal fluid, because with $B^\mu$ order zero in derivatives it already contains all the information related to the anomaly.}

{\ We geometrise the thermodynamic sources as in \eqref{Eq:ThermoFrameDefinition}. Subsequently, we can write down the equilibrium generating functional order by order in derivatives. Our derivative counting is such that hydrostatic constraints, given by acting with $V^{\mu}$ on the metric, gauge field and thermodynamic quantities \eqref{Eq:ThermoFrameDefinition}, are always order one or higher in derivatives. For example,
	\begin{eqnarray}
		&\;& \mathcal{L}_{V}(T) = 0 \; \Rightarrow	\; \frac{\Delta^{\mu \nu} \nabla_{\nu} T}{T} + a^{\mu} = 0 + \mathcal{O}(\partial) \; , \\
		&\;& a^{\mu} = u^{\nu} \nabla_{\nu} u^{\mu} \; . 
	\end{eqnarray}
The hydrostatic constraints give relations between thermodynamic quantities, but the derivative order of the quantities themselves is a priori undetermined. For example, it is easy to imagine situations where the vorticity and the magnetic field are order one or order zero in derivatives. We shall consider the magnetic field at order zero here; all other quantities built from the metric and gauge field and their derivatives with the exception of $T$, $\mu$, $u^{\mu}$ and $B^{\mu}$ we choose to be order one in derivatives.}

{\ The definitions given in \eqref{Eq:ThermoFrameDefinition} specify completely the thermodynamic frame at the hydrostatic level. However, when considering dissipative corrections to the fluid, the frame will be ambiguous up to redefinitions of the chemical potential, temperature and fluid velocity in terms of the hydrostatic constraints. Nevertheless this will not matter for our discussion, as we will be only dealing with the ideal fluid. Despite this, any frame build upon the ideal fluid will not lead to frame-dependent anomalous conductivities.}

{\ At zero order in derivatives the hydrostatic generating functional for the constitutive relations takes the form
	\begin{eqnarray}
		W[g,A,F] = \int d^{3+1}x \sqrt{-g} \; p(T,\mu,B^2)
	\end{eqnarray}
where again $p$ is the pressure. Varying with respect to the background metric and gauge field allows us to identify the one-point functions:
	\begin{eqnarray}
		\delta W &=& \int d^{3+1}x \sqrt{-g} \; \left( \frac{1}{2} T^{\mu \nu} \delta g_{\mu \nu} +J^{\mu} \delta A_{\mu} \right. \nonumber \\
			      &\;& \left. \hphantom{\int d^{3+1}x \sqrt{-g} \; \left( \right.} + \frac{1}{2} M^{\mu \nu} \delta F_{\mu \nu} \right) \; . 
	\end{eqnarray}
Subsequently, employing the decompositions defined in \eqref{Eq:TensorCurrentDecomp} we find
\begin{subequations}\label{eqn:orderzeroconstitutiverelations}
\begin{align}
    \mathcal{E}&=-p+sT+\mu n \\
    \mathcal{P}&=p-\frac{2}{3}\chi_BB^2 \\
    \mathcal{Q}^\mu&=-\chi_B\varepsilon^{\mu\nu\rho\sigma}u_\nu E_\rho B_\sigma+\xi_B^\epsilon B^\mu \\
    \mathcal{T}^{\mu\nu}&=\chi_B\left(B^\mu B^\nu-\frac{1}{3}\Delta^{\mu\nu}B^2\right) \\
    \mathcal{N}&=n \\
    \mathcal{J}^\mu&=\varepsilon^{\mu\nu\rho\sigma}u_\nu\nabla_\rho\mathfrak{m}_\sigma+\varepsilon^{\mu\nu\rho\sigma}u_\nu a_\rho\mathfrak{m}_\sigma+\xi_BB^\mu
\end{align}
where the magnetization is
\begin{eqnarray}
    \mathfrak{m}^\mu =\chi_BB^\mu
\end{eqnarray}
\end{subequations}
and each of the terms now depends on $T$, $\mu$ and $B^2$, while the magnetic susceptibility is defined as $\chi_B=2\frac{\partial p}{\partial B^2}$.}

{\ To obtain the contribution of the anomaly to the constitutive relations we fluctuate the generating functional
\begin{multline}
	W_\text{anom}=\int\mathrm{d}^{3+1}x\sqrt{-g}\ \frac{c}{3}\mu\ B^\mu A_\mu\\
	-\frac{C}{24}\int\mathrm{d}^{4+1}x\sqrt{-G}\varepsilon^{mnopq}A_mF_{no}F_{pq}\ .
\end{multline}
The first term gives the consistent current anomalous transport coefficients, while the second term is a Chern-Simons functional whose fluctuation on the boundary gives the Bardeen-Zumino current that makes the total current covariant. We then find the same transport coefficients that appear in the thermodynamic frame \eqref{Eq:ThermodynamicFrame}\footnote{This is not surprising, since we are considering the same generating functional in the thermodynamic frame, simply pushing $B^\mu$ one order lower in derivative.}
	\begin{equation}
	\xi^\epsilon_B=\frac{1}{2}c\mu^2 \; , \qquad\qquad	\xi_B=c\mu		\ .
\end{equation}
}

{\ Given the constitutive relations, we once again, linearise around an equilibrium with zero velocity, constant temperature, chemical potential and magnetic field. Subsequently,  we find the following expressions for the non-zero frequency longitudinal conductivities
\begin{subequations}\label{eqn:idealconductivities}
\begin{align}
    \sigma(\omega)&= \frac{i}{\omega} \left[ \frac{n^2}{(p+\epsilon)} + \Xi B^2 \right] + \mathcal{O}(\omega^{0}) \; , \\
       \alpha(\omega) &= \frac{i}{\omega} \left[ \frac{ns}{(p+\epsilon)} - \mu \Xi B^2 \right] + \mathcal{O}(\omega^{0}) \; , \\
    \kappa(\omega) &=  \frac{i}{\omega} \left[ \frac{s^2T}{(p+\epsilon)} + \frac{\mu^2 \Xi B^2}{T} \right] + \mathcal{O}(\omega^{0}) \; , 
\end{align}
where
\begin{align}
    \Xi &= \frac{c^2s^2 T^2( \frac{\partial n}{\partial T} \mu - \frac{\partial \epsilon}{\partial T})}{(p+\epsilon)(\frac{\partial\epsilon}{\partial\mu} \frac{\partial n}{\partial T} -  \frac{\partial \epsilon}{\partial T} \frac{\partial n}{\partial\mu}) +B^2c^2\mu^2( \frac{\partial \epsilon}{\partial T} - \frac{\partial n}{\partial T} \mu)} \; ,
\end{align}
\end{subequations}
and all the thermodynamic functions above depend on $(T,\mu,B^2)$.}

{\ These expressions suffer from none of the problems we found previously: they are frame-independent, Onsager reciprocal and furthermore the electric field and chemical potential gradient are equivalent sources (at least for the longitudinal transport, for which the magnetization does not matter). Notice that because $B$ is now order zero, even if the anomalous transport coefficients have the same form they take in the thermodynamic frame, the conductivities looks different, since there is no need to expand $\Xi$ to leading order in $B$, i.e. because we can now consider large values of $B$.}

{\  Results from kinetic theory \cite{abbasi:hydrodynamicexcitations,hidaka:nonlinearresponses,dantas:magnetotransportmultiweyl,gao:chiralanomaly}, thermal QFT \cite{kharzeev:chiralmagnetic,landsteiner:gravitationalanomaly,landsteiner:gravitationalanomalya}, EFT \cite{dubovsky:effectivefield,haehl:effectiveactions,glorioso:globalanomalies} and even Ward identities \cite{jensen:triangleanomalies} also suggest that the thermodynamic frame transport coefficients are preferred (see \cite{landsteiner:notesanomaly} for a review). Here we see that to obtain non-trivial results involving the anomaly from order one constitutive relations we must work $B^\mu$ order zero.}
	
	{\ Even if the anomalous transport coefficients are the same as in the thermodynamic frame, the result in \eqref{eqn:idealconductivities} are -- to the best of our knowledge -- new. Additionally we learn from these expressions \eqref{eqn:idealconductivities} that: $i$) the dependence of the conductivities on $B$ is not simply quadratic as reported in the literature, specifically at larger values of $B$, and this statement does not rely on the fact that the thermodynamics can depend on $B$, as pointed out already in \cite{landsteiner:negativemagnetoresistivity},  furthermore, at order one in derivatives, there are more transport coefficients than just $\sigma_0$ \cite{ammon:chiralhydrodynamics} and $ii$) using purely hydrodynamical arguments (i.e. the requirement that conductivities are frame-independent) we have shown that the magnetic field should be taken to order zero to avoid inconsistencies at all orders in the derivative expansion.

		{\ There is however a small caveat to the above discussion. As alluded to previously the hydrostatic configuration of the system is defined by a Killing vector $V^{\mu}$. Typically, one identifies $V^{\mu} / \sqrt{-V^2} = u^{\mu}$ in the hydrostatic limit and this defines the thermodynamic frame. When the vector field $B^{\mu} \sim \mathcal{O}(\partial^{0})$ however, there is the option to modify this canonical definition e.g. $u^{\mu} = 1/c_{0} \left( V^{\mu} / \sqrt{-V^2} + c_{1} \mu B^{\mu} \right)$ where $c_{0}$ is a normalisation constant so that $u_{\mu} u^{\mu}=-1$ and $c_{1}$ a parameter that can be tuned to set $\mathcal{Q}^{\mu}=0$ - which is part of the definition of the Landau frame \footnote{We thank Ashish Shukla for suggesting this to us.}. However, the resultant expressions look like the Landau frame only if we include the magnetic field up to $\mathcal{O}(B)$ in amplitude i.e. $T^{\mu \nu}  u_{\nu} = - \epsilon u^{\mu} + \mathcal{O}(B^2)$ and $J^{\mu} u_{\mu} = - n + \mathcal{O}(B^2)$. One can of course work with these expressions, although again the Landau frame is out of reach.}
	
\textit{Discussion.}
	{\ It becomes clear from our work that to find anomalous effects in transport with order one magnetic fields one must work to at least order two in derivatives. However, this derivative counting for $B$ would just shift the issue of frame-ambiguities in the conductivities at order $\mathcal{O}(\partial^3)$. To avoid this issue altogether, and indirectly to permit a characterization of transport at larger values of $B$, we suggest one should always work with a magnetic field that is order zero in derivatives, i.e. an integral part of the thermodynamics. Any hydrodynamic frame build upon such an ideal fluid will then give conductivities which are well-defined and not frame-dependent.}
	
\section*{Acknowledgements}

{{\noindent We would like to acknowledge discussions with Pavel Kovtun, Ashish Shukla and Amos Yarom. A.A. and I.M. have been partially supported by the “Curiosity Driven Grant 2020” of the University of Genoa and the INFN Scientific Initiative SFT: “Statistical Field Theory, Low-Dimensional Systems, Integrable Models and Applications”. This project has also received funding from the European Union’s Horizon 2020 research and innovation programme under the Marie Sklodowska-Curie grant agreement No. 101030915. }}

	\bibliography{ref}
	
\end{document}